\documentstyle[prl,aps,preprint]{revtex}

\begin{document}

\preprint{TAUP2403-97}

\title{ Gravitational Instanton for Black Hole Radiation}
\author{S. Massar$^a$ and R. Parentani$^b$}
\address{$^a$ Raymond and Beverly Sackler Faculty of Exact Sciences,\\
School of Physics and Astronomy,
Tel-Aviv University, Tel-Aviv 69978, Israel\\
$^b$ Laboratoire de Math\'ematique et Physique Th\'eorique,
C.N.R.S. UPRES A 6083,\\
Facult\' e des Sciences,
Universit\' e de Tours,
37200 Tours, France}

\maketitle

\begin{abstract}
Hawking radiation is derived from the existence of a 
euclidean
instanton which lives 
in the euclidean 
%continuation of the
black hole geometry. 
Upon taking into account the gravitational field
of the instanton itself, its action is exactly equal to one quarter the
change in the horizon area. This result
also applies to the Schwinger process, the Unruh process, and particle
creation in deSitter space. The implications for horizon 
thermodynamics are discussed.
\end{abstract}
%\pacs{04.70.Dy}
%\begin{multicols}{2}

The thermodynamic properties of black holes were established more than
20 years ago. They are based on the identification of one quarter 
the horizon area
with an entropy\cite{Bek}, the classical laws of black hole
mechanics\cite{BCH} and the process of black hole
evaporation\cite{H}. Since then much work has been devoted to
obtain a better understanding of their origin, their implications,
and the possible consequences of the gravitational back
reaction. Hawking radiation has been derived in several different ways
%(see for instance \cite{GO})
which all rely to some extent on regularity of the field state at the
horizon, either in the Lorentzian space time, or in its euclidean
continuation. In the latter case one obtains the Hartle-Hawking state
characteristic of eternal black holes. The euclidean continuation of
the black hole geometry has also been the subject of much interest
since the work of Gibbons and Hawking\cite{GH} who interpreted it as
the saddle point contribution to the partition function of 
quantum
gravity.

In this Letter we shall show that Hawking radiation can be derived
from the existence of a euclidean instanton.  A similar instanton
can also be associated to the Schwinger process\cite{S} and pair creation in 
deSitter space\cite{GH2}.
%Thus all these particle creation processes result from
%the existence of tunneling orbits.
The instanton is a periodic, static trajectory which lives 
in the euclidean continuation of space time. 
The exponential of minus its action yields the leading 
contribution to 
the probability
to emit a particle. The most interesting aspect of the present
analysis is to include the gravitational
field of the instanton itself to obtain a self consistent euclidean
solution. The Einstein-matter action is then reexpressed in terms
of boundary terms only\cite{BTZ}\cite{HHor}, whereupon 
the action of the self consistent solution is 
equal to one quarter of the {\it change} of the
 black hole, the Rindler, or the deSitter
horizon
area. 
Thus the rate  of particle production is
\begin{equation}
R=Ne^{-S_E} =Ne^{ \Delta A_H /4}
\label{SE}
\end{equation}
($G=\hbar=c=1$) where $N$ is a phase space factor. 
We note that this expression was first obtained in the context of pair 
production of black holes in an electric field\cite{HHR} and of pair
creation
of black holes in de Sitter space\cite{BH}. In these cases one must also
include on the r.h.s. of eq. (\ref{SE}) the area of the horizon of the
produced black holes. It was pointed out in \cite{HHor} that
eq. (\ref{SE}) applies to the Schwinger process. And Eq. (\ref{SE}) was  
derived in the context of black hole evaporation in \cite{KVK}, both in
a semiclassical calculation performed in a Lorenztian
geometry and in the context of superstring theory,
and it was argued that eq. (\ref{SE}) probably takes into
account gravitational backreaction effects neglected in the
background field approximation used by Hawking.

Before turning to the black hole problem, we first 
consider the Schwinger process in Minkowski coordinates to show 
that the instanton action is the leading contribution to the pair
creation process. We then describe it in Rindler
coordinates. 
The mapping to the black hole
problem  will then be straightforward because of the close analogies
between Rindler coordinates in flat space and Schwarzshild coordinates  
in the black hole geometry.

The classical trajectory of a particle of mass $m$ and charge $e$
 in a
constant electric field $E$ is
\begin{equation}
t(s)=t_0 + {1\over a} sh (ams) \quad z(s) = z_0 + {1\over a} ch (ams)
\label{TX}\end{equation}
where $a= e E/m$ is the acceleration.
The corresponding euclidean trajectory is obtained by taking
$s \to is$:
\begin{equation}
t(s) = t_0 + {i \over a} \sin ( ams) \quad z(s) = z_0 + {1\over a} \cos (ams)
\label{TXE}
\end{equation}
This euclidean orbit naturally arises in the 5th time
formalism to derive the Schwinger pair production.
It also appears upon evaluating the solution of the
Klein-Gordon equation in a WKB approximation\cite{PB5}. 
%The tunneling 
%amplitude associated to it gives rise to the Bogoljubov transformation.
%The euclidean orbit (\ref{TXE}) also arises in the 5th time
%formalism to derive the Schwinger pair production\cite{PB5}. 
In second quantization, the probability not to produce pairs 
decreases as $\vert \langle
0_{out}\vert 0_{in}\rangle\vert^2 = e^{-\Gamma TV}$.
In the 5th time
formalism, the rate of 
pair production per unit volume and unit time $\Gamma$ is
expressed as
%\begin{eqnarray}
$\Gamma T V =
Im
\int\!dx\int_0^{\infty}\! {ds \over s}
\int \!{\cal D}x
\ e^{i S(x,s)}$
%\label{5time}
%\end{eqnarray}
where the path integral is over all periodic paths in time $s$
weighted by the
action 
%\begin{equation}
$S=\int\!ds\ {1 \over 4} (\dot t^2 - \dot z ^2) - eA_t \dot t + m^2
$
%\label{S}
%\end{equation}
where $A_t = E z$. In the above we have taken the transverse
momentum (with respect to the electric field) to vanish. It can easily
be taken into account by replacing $m^2$ by $m^2 + k_\perp^2$, and
integrating at the end of the calculation over $k_\perp$, see \cite{PB5}.
The path integral can be evaluated by saddle
point (which for constant $E$ is exact) to yield
%\begin{equation}
$
\Gamma T V = Im
\int\!dx\int_0
^{\infty}\! {ds \over s}
\ {e^{i S_{class}(s) } \over \sqrt{VV(s)}}
$
%\label{Saddle}
%\end{equation}
where $S_{class}$ is the action to complete a closed trajectory
%hence equal to $m^2 s$**
evaluated  on classical periodic
trajectories with period $s$. The integral over proper time is
entirely dominated by the poles of the VanVleck determinant
($VV$) which occur at the
values of $s= i 2\pi n /a m$ ($n=..,-1,0,1,..$) 
corresponding to $n$ windings along the euclidean
trajectory eq. (\ref{TXE}). 
Their classical action is $S =
i\pi n m^2/eE$. Taking also into account the residue of the poles yields
Schwinger's result
$ \Gamma = ({E \over 2 \pi})^{2} \ln ( 1 + e^{- \pi M^2/E})$.  

%The euclidean trajectories
%eq. (\ref{TXE}) can be derived from the  euclidean action $S_E$,
%obtained by taking 
% $s \to is$,
%$t \to it$ in $S$:
%\begin{equation}
%S_E=\int\!ds\ {1 \over 4} (\dot t^2 + \dot z ^2) - eA_t \dot t + m^2
%\label{Seu}
%\end{equation}
%Ca me semble evident et redit

We now consider the same orbits in Rindler coordinates $(\rho , \tau )$
defined by $t=\rho sh \tau$ and $z = \rho ch \tau$ whereupon the metric is
$ds^2 = -\rho^2 d\tau^2 + d \rho^2$ and the EM potential in the boost
invariant gauge is $A_\tau = E \rho ^2 /2$. 
The  euclidean orbits can be derived from the
euclidean action 
$S_E = \int\!ds {1\over 4} (g_{\tau\tau}\dot \tau^2 +
g_{\rho\rho}\dot \rho^2) - e \dot \tau A_\tau  +m^2$
(obtained from the Lorentzian action by taking $s \to is$ and $\tau \to
i\tau$)
where $g_{\tau\tau}=\rho^2$ and $g_{\rho\rho}=1$.
% are the euclidean metric. 
It is convenient to reexpress $S_E$ in a Hamiltonian decomposition as
$S_E=\int\! ds ( \nu \dot \tau + p \dot \rho - H_E)$ where the
euclidean hamiltonian is
\begin{equation}
H_E = g^{\tau\tau}(\nu + e A_\tau)^2 + g^{\rho\rho}p^2 - m^2
\end{equation}

The instanton, that is the static euclidean trajectory corresponding
to a centered euclidean orbit eq. (\ref{TXE}) with $t_0=z_0=0$, is
obtained from the staticity conditions $p=0$, $\dot p = -
\partial_\rho H_E =0$, the mass shell condition $H_E =0$ and the time
evolution
$\dot \tau = \partial_\nu H_E$. 
Explicitly it reads
$\rho = 1/a$, $\tau = eE s$ and has energy $\nu=  m^2/eE$. 
The periodicity of the euclidean solution  results now
from the topology of
the euclidean continuation of Rindler space which 
is a plane described in polar
coordinates. Regularity of the space at the origin imposes that
$\tau$ (the polar coordinate) be periodic with period $2 \pi$.  
The action for one period is 
simply $S_E = \nu \Delta \tau =  \pi m^2/ eE$.
It equals the action selected by the first pole ($n=1$)
of the Van Vleck.

We now introduce the electric and 
gravitational field of the instanton, whereupon we will find that 
$S_E$ is equal to one quarter the change in the area of the
acceleration horizon.  In order to control the self interactions of the
instanton
it is necessary that it no longer describes a point
particle (due to  its gravitational field a point particle
becomes a black hole), but an
extended object. The end result is independent of the nature of the
extended object. For definiteness we take it to be a 
sheet which extends in the transverse direction.
 Inserting this in the
Einstein-Maxwell action -expressed in a Hamiltonian decomposition-
yields
\begin{eqnarray}
S_E&=& \int \!d\tau\left(\int_{\Sigma_\tau} p^{ij}\partial_\tau g_{ij} + E^i 
\partial_\tau A_i + p
  \partial_\tau \rho \right. \nonumber\\
& &\ \left. - N {\cal H} - N^i {\cal H}_i - N A_\tau
({\rm div} E - J^\tau))\right) + S_{Boundary}
\label{Sg}
\end{eqnarray}
where ${\cal H}$ and ${\cal H}_i$ are the hamiltonian and momentum
constraints respectively.
The boundary terms arise because one requires that the variation of
$S_E$ with the physical boundary data fixed yield the equations  of
motion. In the present case the boundary conditions which are kept
fixed are the asymptotic metric and electric field at infinity, 
and the vanishing 
of the laps at the horizon. This latter condition implies that
the boundary term at the horizon is equal to one quarter the 
(infinite) area
of the Rindler horizon.
The euclidean action of the instanton  is obtained by
subtracting from $S_E$ the action without the shell present, but with
the same boundary data at infinity. 
Because of this condition, the staticity of the
instanton and the constraints, the action of the 
instanton is given by 
the (finite) difference of the boundary terms at the horizon only
\begin{equation}
S_E(instanton)= \Delta {A}_H/4 
\label{a/4}
\end{equation}

To  prove that $S_E = \Delta A_H/4$ behaves like an
entropy we must consider a situation of thermal equilibrium. 
We therefore analyze the equilibrium of  a uniformly accelerated detector 
with the Unruh heat bath\cite{U}. The detector model we
use is that of a
 ``two level ion'' in an electric
field. 
By two level ion we mean a
particle of mass $m$ and charge $e$ which can make a transition, by
emitting a massless chargless quantum, to an excited state of rest
mass $m'$. In a constant electric field pairs of ions can be
produced, in either the ground or excited state, with probability
$P_{pair\ m}= e^{- \pi m^2/eE}$ and 
$P_{pair\ m'}= e^{- \pi m'^2/eE}$ respectively. On the other hand
such a two level ion in an electric field behaves like a Unruh
detector since it can make transitions between its internal states
by absorbing or emitting a Rindler quantum. 
In \cite{ShU} it was shown that the rates ($R_{m\to m'}$ and $R_{m' \to
  m}$) for the Unruh process are
related to the probability for the Schwinger process by the exact
relation (obtained by taking into account corrections in $(m'
-m)/m$ and $m/a$):
\begin{equation} 
{ R_{m\to m'} \over R_{m'\to m} }=
 { P_{pair\ m'} \over  P_{pair\ m} }=
 e^{-\left(S_E(m') - S_E(m)\right)}
\label{RP}\end{equation}
(which  can be understood by noting that the amplitudes
describing the Schwinger and Unruh processes 
are related one to another by level
crossing and CPT\cite{ShU}).
Unruh's result, namely that the detector perceives a thermal bath at
temperature $T=a/2\pi$, is recovered in the 
(canonical) limit $m' - m << m$ whereupon 
$S_E(m') - S_E(m) = (m'-m) \partial_m S_E(m)$. 
Indeed direct differentiation yields $\partial_m
S_E(m) = \partial_m  \pi m^2/eE = 2 \pi / a$. 
The origin of this canonical result can  be
understood by recalling that $S_E = \Delta \tau \nu$ and by
expressing $\nu$ as
 $\nu (m, e, \rho) = g_{\tau\tau}(\rho)^{1/2} m - e A_\tau(\rho)$
which results from the mass shell condition $H_E =0$, whereupon 
$\partial_m S_E(m)= \Delta \tau (\partial_m \nu + (d\rho / dm)\partial_\rho
\nu) = \Delta \tau \partial_m \nu  = 
\Delta \tau g_{\tau\tau}^{1/2}$ since $\partial_\rho
\nu =0$ follows from the staticity condition $\partial_\rho H_E
=0$. Thus $\partial_m S_E(m)$ is indeed the proper time to wind once
round
the orbit. 

When the two level ion is in thermal equilibrium with the Unruh heat
bath the probability to find it in the ground or excited state is
given by
$P_{m'} / P_m = { R_{m\to m'} / R_{m'\to m} }$. By virtue of
eq. (\ref{RP}) and eq. (\ref{SE}) this is also equal to
\begin{equation}
{P_{m'} \over P_m} = exp ({A_H(m)/4 -  A_H(m')/4})
\label{PA/4}
\end{equation}
where we have used the fact that 
physical processes are always governed by
area changes to rewrite 
$\Delta  A_H(m)/4 -  \Delta A_H(m')/4 = A_H(m)/4 - A_H(m')/4$.
with $A_H(m^{(')})$ the area of the horizon with a detector of mass
$m^{(')}$ present.  Eq. (\ref{PA/4}) 
is identical to that which would be obtained if the
detector and the horizon form a microcanonical ensemble\cite{ShU}.

We now turn to the black hole problem. In this letter we shall only
exhibit the  
periodic, stationary
orbits living in the euclidean black hole geometry. In a future
publication we intend to report on the relation between these orbits
and more conventional derivations of black hole radiation. 
We first consider the emission of charged
massive particles by a charged black hole. It was shown by
Gibbons\cite{G} that the evaporation of such a black hole is indeed
controlled by the Schwinger process rather than the Hawking
process. Therefore the preceding formalism should apply with minor
modifications to this case. 

The metric for such a charged black hole is $ds^2 = -(1-2M/r
+Q^2/r^2)dt^2 + (1-2M/r
+Q^2/r^2)^{-1}dr^2 +r^2 d\Omega^2$ and the EM potential regular at the
horizon $r=r_+$ is $A_t = Q/ r - Q/r_+$. By making the change of
variables $\rho =2 r_+ \sqrt{r - r_+} / \sqrt{r_+ - r_-}$ the 
metric and EM potential
reduce near the horizon to the flat space problem in Rindler
coordinates. The action which yields the euclidean trajectories,
obtained from the Lorentzian by taking $s \to is$, $t \to it$, is
\begin{eqnarray}
S_E = \int \! ds\  {1\over 4} (g_{tt} \dot t^2 + g_{rr}\dot r^2 )
-e\dot t A_t -m^2
\label{HE}\end{eqnarray}
where $g_{\mu\nu}$ is the euclidean metric.
We have taken the angular
momentum of the particle to vanish. It can easily be reeinstored by
adding to $m^2$ the term $l^2/r^2$. 
The determination of the static euclidean orbit and of its action is
obtained in the hamiltonian formalism exactly as in the electric
case. The action of the euclidean orbit is $S_E = \beta \nu$ where
$\beta$ in the periodicity of the euclidean time and $\nu$ the energy
of the instanton.
If the parameters $m, e, M, Q$ are such that the 
instanton  lies in the region
near the horizon where the correspondence with Rindler space
obtains, then
$S_E(instanton) \simeq \pi m^2/ e E$ with $E \simeq Q/ r_+^2$ 
the electric field at
the horizon, as in the Schwinger process.

Our aim however is to make contact with the Hawking process {\it per
  se}. A possible procedure would be to take the charge $Q$ to
zero in the above problem. In this case the euclidean static orbit
disappears (this difficulty plagued an early attempt
to apply the 5th time formalism to black hole evaporation\cite{PB}). 
In order to keep a static orbit as $Q\to 0$, one can either consider a
particle with angular momentum, whereupon there is 
a static orbit at a finite distance from the horizon
(for massless particles this is the well known orbit at $r=3M$) whose
euclidean action for one period  is once more  $S_E = \beta  \nu$.
Or, for a particle with zero angular momentum one can adopt the following
indirect procedure. We consider
a
Schwarzshild black hole surrounded at a very large distance $r_0$ by a
pair of spherically symmetric condenser plates. The geometry of this
configuration is $
ds^2 =- \gamma(r) (1-2m(r)/r)dt^2 + (1-2m(r)/r)^{-1} dr^2 +r^2 d\Omega^2
$
with $m(r<r_0 - \Delta) = m_{BH}$, $m(r>r_0 + \Delta) = m_{BH} + m_C$
and $\gamma(r>r_0 + \Delta) = 1$,
$\gamma(r<r_0 - \Delta) = O(m_C/ r_0) $  
where $m_{BH}$ is the black hole mass, $m_C$
is the energy of the condenser plates and $\Delta$
is the separation between the plates. The electromagnetic potential
is
$A_t(r<r_0 - \Delta) = 0$, $A_t(r>r_0 + \Delta)= \phi=cst$. 
Moreover the electric field between the condenser plates is
taken to be such that it exactly counterbalances the gravitational
attraction for a particle of mass $m$ and charge $e$:
$E = \partial_r A_t \simeq M m / e r_0^2$. With this configuration such a
particle will have a static orbit between the two condenser plates. 
The euclidean continuation
of this orbit is the instanton for the Hawking process. Its action
$S_E$ tends to $\beta m $ as $r_0 \to \infty$, corresponding to the probability
for a Schwarzshild black hole to emit a particle of mass $m$ which
reaches infinity with zero kinetic energy.

It may seem surprising that this instanton 
is located at $r=\infty$ whereas black hole evaporation is
usually associated with the existence of a horizon. But the existence
of the horizon is crucial in the above calculation since it imposes
the periodicity of the
euclidean geometry in imaginary time.
On the other hand the Hawking temperature, ie. the surface
gravity, is determined by the whole geometry between the horizon and
$r=\infty$. Together these combine to yield for the euclidean
action the value $S_E = \beta m$ expected for Hawking radiation.

The inclusion of the gravitational and electric field of the instanton
proceeds exactly as in eq. (\ref{Sg}). The action 
of the static matter-gravitational configuration can be
expressed
as  boundary terms only. The action of the instanton itself
is obtained by subtracting
 the action of the black hole with
the same mass and charge at infinity, but with the instanton absent.
Whereupon one obtains that $S_E= \Delta A_H/4$. 
The equality of $S_E=\Delta A_H/4$ and $S_E = \beta \nu$ for infinitesimal
$m$ and $e$ 
is a reexpression of the first law of black hole mechanics
$\Delta A_H/4 = \beta \int_\Sigma \Delta T_t^t - A_t \Delta J^t$. This
can be taken as an 
explanation of  why Hawking radiation is consistent with the 
first law.

Particle production in
deSitter space 
can be analyzed in similar manner. DeSitter space in static
coordinates possesses a horizon at $r=\sqrt{3/\Lambda}$.  
 Regularity of the horizon in euclidean
continuation implies that the euclidean time is periodic with period
$\beta = 2 \pi \sqrt{3/\Lambda}$. 
A massive particle possesses
a static, periodic, euclidean orbit at the origin whose 
action is $m\beta$
corresponding to the probability to produce a particle of mass $m$. 
The gravitational field of the instanton can be
included as above and expressed as the change of the area of the
deSitter horizon.

As for the Schwinger process, one can also consider two particles of
neighboring masses which can make transitions from one mass state to
the other thereby behaving like a particle detector. Then, following
eq. (\ref{RP}), 
we conjecture that the
probability for the detector to make a transition is given by the
ratio of the probabilities for the production of the particle in its excited
and ground state
$P_{m' \to m}/ P_{m \to m'}= e^{-\left(S_E(m') - S_E(m)\right)}$. As in the
detector case, $A_H/4$ 
then behaves like the entropy of the macroscopic
system with which the detector is in thermal equilibrium. When the
energy difference 
 between detector states is small then a first order expansion
of the difference $S_E(m) - S_E(m')$ is legitimate and one recovers
the notion of temperature. The inverse temperature is given by
the proper time to make one orbit $\partial_m S_E(m) = \beta
\sqrt{g_{tt}}$ as in the electric case. 

Recently, Jacobson\cite{J}
showed that the identification of horizon area with an entropy and
of inverse acceleration with a temperature, together with the first
law of thermodynamics, imply Einstein's equations. It is most
remarkable that the two ingredients used by Jacobson can be derived from the
(quantum) relation eq. (\ref{SE}). 
It is also important to note the connection between  
eq. (\ref{SE}) and the holographic 
hypothesis\cite{HOL} wherein it is postulated that all the degrees
of freedom inside a given volume are encoded in the surface surrounding
the volume. Here it is the processes which occur inside, or outside,
the volume which are encoded in the surface surrounding it.
These results show the fundamental r\^ole
causal horizons play in the quantum theory of gravity.

We would like to thank 
Y. Aharonov, R. Brout, F. Englert, S. Itzhaki, S. Nussinov,
Ph. Spindel and Y. Zanelli for stimulating discussions.

%\end{multicols}

\end{document}